\documentstyle[prl,aps,preprint,eqsecnum]{revtex} 
\begin{document} 
\draft 
\title{Wave propagation in linear electrodynamics} 
 
\author{Yuri N.\ Obukhov\footnote{Department of Theoretical Physics, 
Moscow State University, 117234 Moscow, Russia. E-mail: 
yo@thp.uni-koeln.de}, 
Tetsuo Fukui\footnote{On leave from: Department of Human Informatics, 
Mukogawa Women's University, 663-8558 Nishinomiya, Japan. E-mail: 
fukui@mwu.mukogawa-u.ac.jp}, 
and Guillermo F. Rubilar\footnote{E-mail: gr@thp.uni-koeln.de}} 
\address 
{Institute for Theoretical Physics, 
University of Cologne\\ 
D-50923 K{\"o}ln, Germany} 

\maketitle

\begin{abstract} 
The Fresnel equation governing the propagation of electromagnetic
waves for the most general linear constitutive law is derived. The
wave normals are found to lie, in general, on a fourth order surface.
When the constitutive coefficients satisfy the so-called reciprocity
or closure relation, one can define a duality operator on the space
of the two-forms. We prove that the closure relation is a sufficient 
condition for the reduction of the fourth order surface to the familiar 
second order light cone structure. We finally study whether 
this condition is also necessary.
\end{abstract} 
\bigskip\bigskip 
\bigskip 
\pagebreak 
 
\section{Introduction} 
 
Electromagnetic wave represents perhaps the most important classical
device with the help of which one can carry out physical measurements
and transmit information. Intrinsic properties and motion of material 
media, as well as the geometrical structure of spacetime, can affect 
the propagation of electromagnetic waves. In the most general setting
\cite{general,axiom}, electromagnetic phenomena are described by the 
pair of two-forms $H, F$ (called the electromagnetic excitation and the 
field strength, respectively) which satisfy the Maxwell equations $dH =J, 
dF =0$, {\it together} with the constitutive law $H = H(F)$. The latter 
relation contains a crucial information about the underlying physical 
continuum (i.e., material medium and/or spacetime). Mathematically, 
this constitutive law arises either from a suitable phenomenological 
theory of a medium or from the electromagnetic field Lagrangian. 

In general, the constitutive law establishes a nonlinear (or even 
nonlocal) relation between the electromagnetic excitation and the 
field strength. The function (or functional) $H(F)$ may depend on the 
polarization and magnetization properties of matter, and/or on the 
spacetime geometry, i.e. metric, curvature, torsion and nonmetricity. 
Previously, propagation of electromagnetic waves was analysed for a 
variety of constitutive laws: for nonlinear models in Minkowski and 
Riemannian spacetimes \cite{pleb}, for electrodynamics in a 
Riemann-Cartan manifold \cite{RCoptics}, and also for certain nonminimal 
and higher derivative gravity models \cite{high}. Numerous authors 
\cite{riem} discussed the electromagnetic waves in the Einstein-Maxwell 
theory. The main aim of this paper is to investigate 
the wave propagation in the Maxwell electrodynamics with the most general 
{\it linear} constitutive law. We derive the generalized Fresnel equation 
which determines the wave normals directly from the constitutive 
coefficients. This result is of interest, e.g., for various applications 
in crystaloptics and related domains.

Another motivation for the present work came from the study of a deep
relationship between the duality operators defined on two-forms and the 
conformal classes of spacetime metrics in four dimensions. Within the
classical Maxwell electrodynamics, Toupin, Sch\"onberg and others
\cite{recipr} have noticed that the constitutive coefficients define
a duality operator, provided a certain reciprocity or closure condition 
is fulfilled, and gave first demonstrations of the existence of the 
corresponding conformal metric structure. Later these observations were 
rediscovered and developed in mathematics \cite{dualop1} and in the
gravity theory \cite{dualop2}. Recently the complete explicit solution
of the closure relation has been given \cite{OH}, and it was conjectured
that the reciprocity condition is a necessary and sufficient condition
for the standard null-cone structure for the light propagation (see 
also independent arguments in \cite{Lam99}). Here we give a partial 
answer to this question.

\section{Electrodynamics with linear constitutive law}

Let us consider the Maxwell equations in vacuum, 
\begin{equation} 
dH = 0,\qquad dF =0,\label{dhf} 
\end{equation} 
i.e. we assume that the electric current three-form $J$ vanishes
in the spacetime region under consideration. Given the local 
coordinates $x^i$, $i=0,1,2,3$, we can decompose the exterior forms as
\begin{equation} 
H = {\frac 1 2}\,H_{ij}\,dx^i\wedge dx^j,\qquad 
F = {\frac 1 2}\,F_{ij}\,dx^i\wedge dx^j.\label{geo1} 
\end{equation}
Following \cite{OH,HOR}, we write the linear constitutive law in terms 
of the electromagnetic excitation and field strength tensors as
\begin{equation}\label{cl} 
H_{ij}=\frac{1}{4}\,\epsilon_{ijkl}\,\chi^{klmn}\,F_{mn},\qquad
i,j,\dots =0,1,2,3.
\end{equation} 
Here $\epsilon_{ijkl}$ is the Levi-Civita symbol and $\chi^{ijkl}(x)$ an 
even tensor density of weight $+1$ (called the constitutive tensor density)
which can be decomposed according to
\begin{equation}
{\chi}^{ijkl}=f(x)\,\stackrel{\rm o}{\chi}{}^{ijkl} +\alpha(x)\,
\epsilon^{ijkl} \,,\qquad{\rm with}\qquad \stackrel{\rm o}{\chi}
{}^{[ijkl]}\equiv 0\,.\label{decomp}
\end{equation} 
Here $f(x)$ is a dimensionfull scalar function such that
$\stackrel{\rm o}\chi{}^{ijkl}$ is dimensionless. The pseudo-scalar
constitutive function $\alpha(x)$ can be identified (on the 
{\it kinematic} level) as an Abelian axion field, whereas $f(x)$
can be interpreted as a dilaton scalar field. 
Note that ${\stackrel{\rm o}{\chi}}{}^{ijkl}$ has the same algebraic 
symmetries and therefore the same number of 20 independent components 
as a Riemannian curvature tensor:
\begin{equation}
\stackrel{\rm o}{\chi}{}^{ijkl}= -\stackrel{\rm o}{\chi}{}^{jikl}=
-\stackrel{\rm o}{\chi}{}^{ijlk}= \ \stackrel{\rm o}{\chi}{}^{klij}\,,
\qquad \stackrel{\rm o}{\chi}{}^{[ijkl]}=0\,.\label{symmetries}
\end{equation}
This follows from the existence and the structure of the Lagrangian for 
the linear electrodynamics $V_{\rm lin} = -\,{\frac 1 2}\,H\wedge F$,
see \cite{axiom,HOR}. It is convenient to adopt a more compact (essentially 
{\it bivector}) notation by defining the 3-(co)vector quantities
\begin{equation}\label{3vect}
{\cal D}^a := \left(\begin{array}{c}H_{23}\\ H_{31}\\ H_{12}\end{array}\right), 
\quad 
{\cal H}_a := \left(\begin{array}{c}H_{01}\\ H_{02}\\ H_{03}\end{array}\right),
\qquad {\rm and}\qquad
B^a := \left(\begin{array}{c}F_{23}\\ F_{31}\\ F_{12}\end{array}\right), 
\quad 
E_a := \left(\begin{array}{c}F_{10}\\ F_{20}\\ F_{30}\end{array}\right), 
\end{equation} 
for the electric and magnetic excitations, and for the magnetic and 
electric field strengths, respectively. The Latin indices label now
$a,b,c,\dots = 1,2,3$. The constitutive tensor is then naturally 
parametrized by a triplet of $3\times 3$ matrices, 
${\stackrel{\rm o}{\chi}}{}^{ijkl} = \left\{{\cal A}^{ab}, {\cal B}_{ab},
{\cal C}^a{}_b\right\}$, so that the constitutive law (\ref{decomp}) is 
finally recasted into
\begin{equation} 
\left(\begin{array}{c} {\cal H}_a \\ {\cal D}^a\end{array}\right) = 
f(x)\left(\begin{array}{cc} {\cal C}^b{}_a & {\cal B}_{ab} 
\\ {\cal A}^{ab}& {\cal C}^a{}_b \end{array}\right)
\left(\begin{array}{c} -E_b\\ B^b\end{array}\right) + 
\alpha(x)\left(\begin{array}{c} -E_a \\ B^a\end{array}\right).\label{CR} 
\end{equation}
Here the $3\times 3$ matrices satisfy ${\cal A}^{ab}={\cal A}^{ba}$, 
${\cal B}_{ab}={\cal B}_{ba}$, and ${\cal C}^a{}_a = 0$, thereby 
providing the algebraic properties (\ref{symmetries}).

\section{Wave propagation: Fresnel equation}

In the theory of partial differential equations, the propagation of 
waves is described by Hadamard discontinuities of solutions across a 
characteristic (wave front) hypersurface $S$ \cite{Had}. One can locally 
define $S$ by the equation $\Phi(x^i) = const$. The Hadamard discontinuity 
of any function ${\cal F}(x)$ across the hypersurface $S$ is determined 
as the difference $\left[{\cal F}\right](x) := {\cal F}(x_+) - 
{\cal F}(x_-)$, where $x_{\pm}:=\lim\limits_{\varepsilon\rightarrow 0}
\,(x\pm\varepsilon)$ are points on the opposite sides of $S\ni x$. 
An ordinary electromagnetic wave is a solution of the Maxwell equations 
(\ref{dhf}) for which the derivatives of $H$ and $F$ have regular 
discontinuities across the wave front hypersurface $S$. 

In terms of the (co)vector components, we have on the characteristic
hypersurface $S$:
\begin{eqnarray}
&& [{\cal D}^a] = 0,\quad [\partial_i{\cal D}^a] = d^a\,q_i,\qquad
[{\cal H}_a] = 0,\quad [\partial_i{\cal H}_a] = h_a\,q_i,\label{had1}\\
&& [B^a] = 0,\quad [\partial_iB^a] = b^a\,q_i,\qquad
[E_a] = 0,\quad [\partial_iE_a] = e_a\,q_i,\label{had2}
\end{eqnarray}
where $d^a, h_a, b^a, e_a$ describe discontinuities of the corresponding
quantities across $S$, and the wave-covector normal to the front is given by 
\begin{equation} 
q_i := \partial_i \Phi\,. 
\end{equation} 
Equations (\ref{had1})-(\ref{had2}) represent the Hadamard geometrical
compatibility conditions. Substituting (\ref{geo1}) into (\ref{dhf}), and
using (\ref{3vect}) and (\ref{had1})-(\ref{had2}), we find 
\begin{eqnarray}
&& q_0\,d^a - \epsilon^{abc}\,q_b\,h_c =0,\qquad 
q_0\,b^a + \epsilon^{abc}\,q_b\,e_c =0,\label{geo2}\\ 
&& q_a\,d^a =0,\qquad q_a\,b^a =0,\label{geo3} 
\end{eqnarray}
where $\epsilon^{abc}$ is the 3-dimensional Levi-Civita symbol.
In this system only the 6 equations (\ref{geo2}) are independent. 
Assuming that $q_0\neq 0$, one finds that the equations (\ref{geo3})
are trivially satisfied if one substitutes (\ref{geo2}) into them. 
[Note that the characteristics with $q_0 = 0$ do not have intrinsic
meaning for the evolution equations, since they obviously 
depend on the arbitrary choice of coordinates].
 
Differentiating (\ref{CR}) and using the compatibility conditions
(\ref{had1})-(\ref{had2}), we find additionally 6 algebraic equations: 
\begin{equation} 
\left(\begin{array}{c} h_a \\ d^a\end{array}\right) = 
f(x)\left(\begin{array}{cc} {\cal C}^b{}_a & {\cal B}_{ab} 
\\ {\cal A}^{ab}& {\cal C}^a{}_b \end{array}\right)
\left(\begin{array}{c} -e_b\\ b^b\end{array}\right) + 
\alpha(x)\left(\begin{array}{c} -e_a \\ b^a\end{array}\right).\label{CR1} 
\end{equation}
Note that the constitutive coefficients and their first derivatives 
are assumed to be continuous across $S$. 

We can now substitute $d^a$ and $h_a$ from (\ref{CR1}) into 
the first equation (\ref{geo2}), which gives
\begin{equation} 
f(x)q_0\left(- {\cal A}^{ab}\,e_b + {\cal C}^a{}_b\,b^b\right) + 
\alpha(x)q_0\,b^a = f(x)\epsilon^{abc}q_b\left(- {\cal C}^d{}_c\,e_d + 
{\cal B}_{cd}\,b^d\right) - \alpha(x)\epsilon^{abc}q_b\,e_c. 
\end{equation} 
The terms proportional to the axion field $\alpha(x)$ drop out completely 
due to (\ref{geo2}), and then one can also remove the common dilaton factor 
$f(x)$ on both sides of the equation. [We assume $f(x) \neq 0$, since 
otherwise there is no hyperbolic evolution system]. It remains finally to 
substitute $b^a$ in terms of $e_b$ from the second equation (\ref{geo2}), 
and after some rearrangements one finds: 
\begin{equation}
\left(q_0^2\,{\cal A}^{ab} + q_0q_d\left[{\cal C}^a{}_c\,\epsilon^{cdb} +
{\cal C}^b{}_c\,\epsilon^{cda}\right] + q_eq_f\,\epsilon^{aec}\epsilon^{bfd}
\,{\cal B}_{cd}\right)e_b =0.\label{algeq}
\end{equation}
This homogeneous algebraic equation has a nontrivial solution when
\begin{equation}
{\cal W} := \det\left|q_0^2\,{\cal A}^{ab} + q_0q_d\left[{\cal C}^a{}_c
\,\epsilon^{cdb} + {\cal C}^b{}_c\,\epsilon^{cda}\right] + q_eq_f\,
\epsilon^{aec}\epsilon^{bfd}\,{\cal B}_{cd}\right| =0.\label{deteq}
\end{equation} 
This is a Fresnel equation which is central in the wave propagation
analysis. It determines the geometry of the wave normals in terms of 
the constitutive coefficients ${\cal A}, {\cal B}, {\cal C}$. 
A direct calculation yields the general result: 
\begin{equation} 
{\cal W} = q^2_0 \left( q_0^4 M + q_0^3q_a\,M^a + q_0^2q_a q_b\,M^{ab} + 
q_0q_a q_b q_c\,M^{abc} + q_a q_b q_c q_d\,M^{abcd}\right) =0,\label{detfin} 
\end{equation}
where we have denoted 
\begin{eqnarray}
&& M:=\det{\cal A},\qquad M^a := 2\epsilon_{bcd} 
\,{\cal A}^{ab}{\cal C}^c{}_e\,{\cal A}^{ed},\label{ma1}\\
&& M^{ab}:= {\cal B}_{cd}({\cal A}^{ab}\,{\cal A}^{cd} -
{\cal A}^{ac}\,{\cal A}^{bd})
- \,{\cal A}^{cd}\,{\cal C}^a{}_c\,{\cal C}^b{}_d
+ 4{\cal A}^{ac}\,{\cal C}^b{}_d\,{\cal C}^d{}_c
- 2{\cal A}^{ab}\,{\cal C}^c{}_d\,{\cal C}^d{}_c,\label{ma2}\\
&& M^{abc} := 2\epsilon^{cde}\left[{\cal B}_{df}(
{\cal A}^{ab}\,{\cal C}^f{}_e - {\cal A}^{af}\,{\cal C}^b{}_e)
+ {\cal C}^a{}_e\,{\cal C}^b{}_f\,{\cal C}^f{}_d\right]\label{ma3}\\
&& M^{abcd} := \epsilon^{cef}\epsilon^{dgh}\,{\cal B}_{fh}
\left[\hbox{$\scriptstyle{\frac 1 2}$} \,{\cal A}^{ab}\,{\cal B}_{eg}
- {\cal C}^a{}_e\,{\cal C}^b{}_g\right].\label{ma4}
\end{eqnarray}
Note that only the completely symmetric parts $M^{(a_1\dots a_p)}$, 
$p=2,3,4$, contribute to the Fresnel equation. 
Since $q_0\neq 0$, one can delete the first factor in (\ref{detfin}), 
and thus we finally find that the wave covector $q_i$ lies, in general, 
on a {\it 4th order} surface. This is different from the light {\it cone} 
(i.e., 2nd order) structure which arises only in a particular case. 
In the next section we demonstrate that the latter corresponds to the 
closure condition. Earlier, the relation between the fourth- and the
second-order wave geometry was studied by Tamm \cite{tamm} for a 
special case of the linear constitutive law.

\section{The closure relation as a sufficient condition}\label{sufficient}

The linear constitutive law defines a duality operator when the 
constitutive coefficients satisfy the `reciprocity' or `closure' 
relation \cite{recipr,OH}:
\begin{equation}
{\frac 1 4}\,\epsilon_{ijmn}\epsilon_{pqrs}{\stackrel{\rm o}{\chi}}
{}^{mnpq}{\stackrel{\rm o}{\chi}}{}^{rskl} = - \delta^{kl}_{ij},\label{rec1}
\end{equation}
or in terms of the $3\times 3$ matrices:
\begin{equation}
{\cal A}^{ac}{\cal B}_{cb} + {\cal C}^a{}_c{\cal C}^c{}_b = 
- \delta^a_b, \quad {\cal C}^{(a}{}_c{\cal A}^{b)c} =0,\quad 
{\cal C}^c{}_{(a}{\cal B}_{b)c} =0.\label{rec2}
\end{equation}
The general solution of the closure condition (\ref{rec1})-(\ref{rec2})
reads \cite{OH}:
\begin{eqnarray} 
{\cal A}^{ab} &=& 
{\frac 1{\det{\cal B}}}\,(k^2\,{\cal B}^{ab} - k^a\,k^b) 
- {\cal B}^{ab},\label{Aab}\\ 
{\cal C}^a{}_b &=& {\cal B}^{ad}\,\epsilon_{dbc}\,k^c = 
{\frac 1 {\det{\cal B}}}\,\epsilon^{adc}\,{\cal B}_{db}\,k_c.\label{Cab} 
\end{eqnarray}
Here $k^a$ is an arbitrary 3-vector, $k_b := {\cal B}_{ab}k^a$, 
$k^2 := {\cal B}_{ab}k^ak^b$, and ${\cal B}^{ab}$ denotes the 
inverse matrix to ${\cal B}_{ab}$.
 
Starting from (\ref{Aab})-(\ref{Cab}), the direct calculation yields: 
\begin{eqnarray} 
&& M =-\,{\frac 1 {\det{\cal B}}}\left(1 - 
{\frac {k^2}{\det{\cal B}}}\right)^2,\label{r1}\\ 
&& M^a = {\frac 1 {\det{\cal B}}}\,4k^a\left(1 - {\frac {k^2}{\det{\cal B}}}
\right), \label{r2}\\ 
&& M^{ab} = -\,{\frac 1 {\det{\cal B}}}\,4k^ak^b + 2{\cal B}^{ab} 
\left(1 - {\frac {k^2}{\det{\cal B}}}\right), \label{r3}\\ 
&& M^{abc} = -\,4\,{\cal B}^{b(a}\,k^{c)}, \label{r4}\\ 
&& M^{(abcd)} = -\,(\det{\cal B})\,{\cal B}^{(ab}{\cal B}^{cd)}.\label{r5} 
\end{eqnarray}
Substituting all this into the general Fresnel equation (\ref{detfin}),
we find 
\begin{eqnarray} 
{\cal W}&=& -\,\sigma\,q_0^2\left[{\frac {q_0^2} {\sqrt{|\det{\cal B}}|}} 
\left(1 - {\frac {k^2}{\det{\cal B}}}\right) - 
{\frac {2q_0(q_ak^a)} {\sqrt{|\det{\cal B}}|}} - 
\sqrt{|\det{\cal B}|}\,(q_aq_b{\cal B}^{ab})\right]^2\nonumber\\ 
&=& -\,\sigma\,q_0^2\left(q_iq_j\,g^{ij}\right)^2.
\end{eqnarray} 
Here $\sigma = sign(\det{\cal B})$, and $g^{ij}$ is the (inverse) 
4-dimensional metric which arises from the duality operator and 
the closure relation \cite{OH,HOR}: 
\begin{eqnarray} 
g^{00}&=& {\frac 1 {\sqrt{|\det{\cal B}}|}}
\left(1 - {\frac {k^2}{\det{\cal B}}}\right),\\ 
g^{0a}&=& -\,\frac{k^a}{\sqrt{|\det{\cal B}|}},\\ 
g^{ab}&=& -\,\sqrt{|\det{\cal B}|}\,{\cal B}^{ab}. 
\end{eqnarray} 
This metric $g_{ij}$ (defined up to a conformal factor) always has the 
Lorentzian signature, although it is not necessarily interpretable as 
a spacetime metric (this is a so called {\it optical metric}, in general; 
see, e.g., \cite{kremer}). As shown in \cite{HOR}, the constitutive 
tensor density (\ref{decomp}) can be rewritten in terms of this metric as
\begin{equation}
{\chi}^{ijkl}=f(x)\,\sqrt{-g}\left(g^{ik}g^{jl}
- g^{jk}g^{il}\right) + \alpha(x)\,\epsilon^{ijkl}\,.
\end{equation}

Thus we indeed recover the null cone $q_i q^i = q_iq_j\,g^{ij} =0$ 
structure for the propagation of electromagnetic waves from our general 
analysis: provided the constitutive matrices satisfy the closure 
relation (\ref{rec1})-(\ref{rec2}), the quartic surface (\ref{detfin}) 
degenerates to the null cone for the induced metric $g_{ij}$. 

It is worthwhile to note that the Fresnel equation (\ref{detfin}) can
be rewritten in an explicitly covariant form
\begin{equation} \label{Fresnel}  
G^{ijkl}q_i q_j q_k q_l = 0, \qquad i,j,\dots=0,1,2,3,  
\end{equation}
where the fourth order totally symmetric tensor density $G^{ijkl}$ is 
constructed as the cubic polynomial of the components of the constitutive 
tensor:
\begin{equation}\label{G4}  
G^{ijkl}:=\frac{1}{4!}\ {\stackrel{\rm o}{\chi}}{}^{mnp(i}\,
{\stackrel{\rm o}{\chi}}{}^{j|qr|k}\,{\stackrel{\rm o}{\chi}}{}^{l)stu}  
\,\epsilon_{mnrs}\epsilon_{pqtu}.
\end{equation}
[Here the total symmetrization is extended only over the four indices 
$i,j,k,l$ with all the summation indices excluded]. Tamm \cite{tamm} has 
introduced a similar `fourth-order metric' for the particular case of 
the linear constitutive law.

\section{The closure relation as a necessary condition}

It was conjectured \cite{OH,HOR} that the closure relation is not
only sufficient, but also a necessary condition for the reduction
of the quartic geometry (\ref{detfin}) to the null cone. The complete 
proof of this conjecture requires a rather lengthy algebra and will be
considered elsewhere. Here we demonstrate the validity of the 
necessary condition in a particular case when the matrix ${\cal C}=0$.
 
Putting ${\cal C}^a{}_b=0$, we find from (\ref{ma1})-(\ref{ma4}) that 
$M^a = 0$ and $M^{abc} =0$, whereas 
\begin{eqnarray} 
M^{ab}&=& {\cal B}_{cd}({\cal A}^{ab}\,{\cal A}^{cd} - 
{\cal A}^{ac}\,{\cal A}^{bd}),\label{mab}\\ 
M^{(abcd)}&=& (\det{\cal B})\,{\cal A}^{(ab}\,{\cal B}^{cd)}.\label{mabcd} 
\end{eqnarray} 
Consequently, (\ref{detfin}) reduces to 
\begin{equation} 
{\cal W} = q_0^2\left(\det{\cal A}\,q_0^4 + q_0^2\,\gamma + 
\det{\cal B}\,\alpha\,\beta\right),
\end{equation} 
where  $\alpha:= A^{ab}q_aq_b$, $\beta:= B^{ab}q_aq_b$, and 
$\gamma:= M^{ab}q_aq_b$. Assuming that the last equation describes 
a null cone, one concludes that the roots for $q_0^2$ should 
coincide and thus necessarily
\begin{equation} 
\gamma^2 = 4\det{\cal A}\det{\cal B}\,\alpha\,\beta.\label{abg} 
\end{equation} 
Let us write $(\det{\cal A}\det{\cal B}) = s\,|\det{\cal A}\det{\cal B}|$,
with $s = sign(\det{\cal A}\det{\cal B})$. Then (\ref{abg}) yields 
\begin{equation}
2\sqrt{|\det{\cal A}\det{\cal B}|}\,{\frac\alpha\gamma}= s\,\lambda,\qquad 
2\sqrt{|\det{\cal A}\det{\cal B}|}\,{\frac\beta\gamma}= {\frac 1 \lambda}, 
\end{equation} 
where $\lambda$ is an arbitrary scalar factor. Recalling the definitions 
of $\alpha, \beta, \gamma$, we then find 
\begin{equation}
{\cal A}^{ab} = s\,\lambda^2\,{\cal B}^{ab}.\label{AB} 
\end{equation} 
Consequently, $M = \det{\cal A} = s\,\lambda^6/\det{\cal B}$ and 
$M^{ab} = 2\lambda^4{\cal B}^{ab}$, and therefore one verifies that
\begin{equation}\label{quad}
{\cal W} = {\frac {s\,\lambda^2q_0^2}{\det{\cal B}}}\left(
\lambda^2q_0^2 + s\,q_aq_b\,{\cal B}^{ab}\det{\cal B}\right)^2.
\end{equation}
We immediately see that for $s=-1$ the quadratic form in (\ref{quad}) 
can have either the $(+ - - -)$ signature, or $(+ + + -)$. Similarly, for 
$s=1$ the signature is either $(+ + + +)$, or $(+ + - -)$. Therefore, 
the Fresnel equation describes a correct light {\em cone} (hyperbolic) 
structure only in the case $s=-1$.

Finally, one can verify that the above solutions satisfy
\begin{equation}
{\frac 1 4}\,\epsilon_{ijmn}\epsilon_{pqrs}{\stackrel{\rm o}{\chi}}
{}^{mnpq}{\stackrel{\rm o}{\chi}}{}^{rskl} = s\lambda^2 \delta^{kl}_{ij},
\end{equation}
which for $s = -1$ reproduces the closure relation (\ref{rec1}) after 
a trivial rescaling of the constitutive tensor density (and subsequently 
absorbing the factor $\lambda$ into the `dilaton' field $f$).

\section{Conclusions}

In this paper we have derived, extending the earlier results (see 
e.g., \cite{riem,kremer,tamm}), the Fresnel equation governing 
the propagation of electromagnetic waves for the most general linear 
constitutive law. The wave covector lies, in general, on a {\it fourth 
order surface}. Such generic fourth order structure is not affected
by the axion- and dilaton-like parts of the constitutive tensor. 
Note however that the linear constitutive law $H = \alpha(x) F$ does not lead to 
hyperbolic evolution equations, and hence necessarily $f(x) \neq 0$.

We have proved that the closure relation (\ref{rec1}) is a sufficient
condition for the reduction of the fourth order surface to the familiar
second order light cone structure. The corresponding family of conformally 
related metrics $g$ coincides with that derived in \cite{OH}, see also 
\cite{HOR}. This result may be considered as an alternative (as compared
to Urbantke's scheme \cite{dualop1,dualop2}) derivation of the Lorentzian
metric $g$ from a duality operator. In terms of the Lagrangian, the closure
relation is equivalent to the statement that $V_{\rm lin} = - \,{\frac 1 2}
\,\left[f(x)\,F\wedge{}^\ast F + \alpha(x)\,F\wedge F\right]$, where the
Hodge operator ${}^\ast$ is defined by the metric $g$.

For the special case ${\cal C}^a{}_b=0$ we have proved that the 
requirement of reduction of the fourth order Fresnel structure to a 
second order one implies a relation between the constitutive coefficients 
which is slightly weaker than the closure relation (\ref{rec1}), in that 
it allows for an arbitrary scalar factor. The latter though can be removed 
by the redefinition of the dilaton field $f(x)$. Also the signature of 
the resulting quadratic form is not fixed, so that one has to impose 
hyperbolicity as a separate condition.

It is worthwhile to note that the results obtained can be directly
applied to the refinement and generalization of the previous analyses
of the observational tests of the equivalence principle. See, for 
instance, \cite{equiv}, where some particular cases of the Fresnel
equation have been studied in this context. 

\bigskip
\noindent {\bf Acknowledgments}
\bigskip

We are grateful to Friedrich W. Hehl for useful comments and 
discussion of the results obtained. TF thanks the Institute for 
Theoretical Physics, University of Cologne, for the warm hospitality.
GFR would like to thank the German Academic Exchange Service (DAAD) 
for a graduate fellowship (Kennziffer A/98/00829).


\end{document}